\documentstyle[12pt]{article}
\input{epsf}

\def\a{\alpha}
\def\b{\beta}
\def\m{\mu}
\def\n{\nu}
\def\l{2\pi\alpha'}
\def\d{\partial}

\def\vg{,}
\def\pt{.}
\def\({\left(}
\def\){\right)}
\def\[{\left[}
\def\]{\right]}

\def\beq{\begin{equation}}
\def\eeq{\end{equation}}
\def\bea{\begin{eqnarray}}
\def\eea{\end{eqnarray}}
\def\bq{\begin{quote}}
\def\eq{\end{quote}}

\def\gappeq{\mathrel{\rlap {\raise.5ex\hbox{$>$}}
{\lower.5ex\hbox{$\sim$}}}}

\def\lappeq{\mathrel{\rlap{\raise.5ex\hbox{$<$}}
{\lower.5ex\hbox{$\sim$}}}}

\def\Toprel#1\over#2{\mathrel{\mathop{#2}\limits^{#1}}}

\begin{document}
\pagestyle{empty}
\begin{flushright}
{CERN-TH/2002-163}\\ 
{\ttfamily hep-th/0207211}\\ 
\end{flushright}
\vspace*{5mm}
\begin{center}
{ \Huge  Derivative Corrections\\
 \vspace*{3mm}from Boundary State Computations} \\
\vspace*{19mm}
{\large Pascal Grange} \\
\vspace{0.5cm}

 {\it CERN Theory Division, CH-1211 Geneva 23, Switzerland.}\\
\vspace{0.1cm} 
{\tt{ Pascal.Grange@cern.ch}}\\
\vspace*{7cm}
{\bf ABSTRACT} \\ \end{center}
\vspace*{5mm}
\noindent
 The boundary state formalism is used to confirm predictions from non-commutativity for the derivative corrections to the Dirac--Born--Infeld and Chern--Simons actions, at all orders in derivatives. As anticipated by S. Mukhi, the method applies by induction to every coupling in the Chern--Simons action. It is also used to derive  the  corrections to the Dirac--Born--Infeld action at quadratic order in the field strength.

\vspace*{1cm}
\noindent

\begin{flushleft} 
July 2002
\end{flushleft}

\vfill\eject

\setcounter{page}{1}
\pagestyle{plain}

\section{Introduction}

Non-commutative field theory arose from string theory {\mbox{~\cite{CH, SW}}} in the sense that it was {\mbox{realized that}} a D-brane, in a background with flat metric $g$ and constant $B$-field, inherited {\mbox{non-commutative}} coordinates from the algebra of the coordinates of the end-points of open strings. This algebra generates the Moyal algebra, endowed with the $\star$ product:

$$[X^i,X^j]=i\,\theta^{ij}\,,\;\;\;\;\;\;\theta=B^{-1}\,,$$
$$f\star g=\exp\(\frac{i}{2}\d_i\,\theta^{ij}\,\d'_j\)f(x)g(x')|_{x'=x}\pt$$

\noindent{This holds provided the}  Seiberg--Witten limit is taken, which amounts to $\alpha'\rightarrow 0$ together with a {\mbox{scaling}} of the metric, $g_{ij}\sim\alpha'^2$, with fixed open string parameters. Considering gauge fields on the world-volume then requires regularization of the gauge theory. Depending on the choice of the regularization scheme, one obtains either ordinary Abelian gauge theory or non-commutative Yang--Mills theory. In order for the field theory not to depend on the regularization scheme, there must be a field redefinition between gauge fields, known as the Seiberg--Witten map ~\cite{SW}, which transforms commutative fields into non-commutative fields. The form of this map was conjectured by Liu in ~\cite{L}.\\

This argument from non-commutative field theory on the world-volume of the brane was combined with string spectra. Since there are two descriptions of the same system using two different geometries, there must be two different expressions for the coupling to a Fourier mode of, say, each of the Ramond--Ramond fields. This led ~\cite{CSterms, LM, OO} in the Abelian case to the derivation of an expression for the Seiberg--Witten map for the field strength, which was shown explicitly to satisfy all the requirements (Bianchi identity, consistency with gauge invariance in the two descriptions, initial condition from vanishing non-commutativity).\\

Furthermore, it is worth looking for derivative corrections to the D-brane action ~\cite{AT} ; small field strengths are not necessarily slowly varying. Once the map between gauge fields is spe\mbox{cified}, the remainder of the action is sufficiently constrained to yield insights on the derivative corrections to the gauge sector. This path was followed by Das, Mukhi and Suryanarayana in ~\cite{DMS}. The {\mbox {Seiberg--Witten}} limit ensures that the corrections to the non-commutative action are suppressed by powers of $\alpha'$. In this limit the non-commutative action is therefore exactly known, once ~\cite{CSterms, Gross} a straight open Wilson line $W_k$, associated with some fixed momentum $k$, is inserted to ensure gauge invariance. The requirement that the commutative and non-commutative actions $S$ and $\hat{S}$ be the same leads to predictions for the derivative corrections in the commutative description. Throughout this work we will consider a Euclidean D9-brane in type IIB string theory. The results hold for flat lower-dimensional D-branes. The {\mbox{Ramond--Ramond}} form fields  $C^{(2p)}$, with $0\leq p\leq 5$, will be collectively denoted by $C$. Treating separately the coupling to a mode $\tilde{D}(-k)$ of the dilaton, or Dirac--Born--Infeld action, 
and the coupling to a mode $\tilde{C}(-k)$ of the Ramond--Ramond form fields, or Chern--Simons action, the predictions in the Seiberg--Witten limit involve derivatives of all orders. They are encoded in the following formulas:

$$S_{DBI}+\Delta S_{DBI}= \hat{S}_{DBI}= \frac{\tilde{D}(-k)}{g_s}\int d^{10}x\, L_\star \(\frac{{\rm Pf} Q}{{\rm Pf} \theta}\sqrt{\det(g+\l Q^{-1})}W_k(x)\)\star e^{ikx}\vg$$
$$S_{CS}+\Delta S_{CS}= \hat{S}_{CS}=\tilde{C}(-k)\wedge \int L_\star \(\frac{{\rm Pf} Q}{{\rm Pf} \theta} e^{\l Q^{-1}} W_k(x) \)\star e^{ikx}\pt$$

 These expressions are written in terms of non-commutative fields, and use is made of the tensor $Q^{\m\n}=\theta^{\m\n}-\theta^{\m\rho}\hat{F}_{\rho\tau}\theta^{\tau\n}$, the field $\hat{F}$ being the non-commutative counterpart of the field strength $F$. In terms of the commutative variables, one obtains for the corrected Chern--Simons action:

$$S_{CS}+\Delta S_{CS}=\int C\wedge \sum_{p}\frac{1}{p!}\star_p[F^p]\pt$$

Apart from the prefactors, and the integration $\int d^{10}x\frac{\sqrt{(g+\l B)}}{g_s}\(\dots\)$, one is led to the following for the terms of order $F^2$ in the corrected Dirac--Born--Infeld action:

$$ S_{DBI}+\Delta S_{DBI}=\frac{1}{4}\theta^{\m\n}\theta^{\rho\tau}\star_2[{F}_{\n\rho}, {F}_{\tau\mu}]-\frac{1}{8}\theta^{\m\n}\theta^{\rho\tau}\star_2[{F}_{\m\n}, {F}_{\rho\tau}]\vg$$

\noindent{where} the $\star_p$ products are in both cases reminiscent of the $L_\star$ integration prescription of $p$ operators along the Wilson line. Explicit forms for these commutative but non-associative products, mapping $p$ functions to one function, were derived in ~\cite{L}. For example $\star_2$ is expressed as the following series:

$$\star_2[f(x),g(x)]=\frac{\sin(\frac{\d_\lambda\theta^{\lambda\rho}\d'_{\rho}}{2})}{\frac{\d_\lambda\theta^{\lambda\rho}\d'_{\rho}}{2}}f(x)g(x')|_{x'=x}\pt$$

As suggested by a comparison of the first orders in derivatives with the terms computed by Wyllard in ~\cite{W} using the boundary state formalism, these results could well be recovered through computations on an ordinary {\it commutative} space. Successful checks were made at finite order in ~\cite{DMS}; more precise ones, within the framework of the {\mbox{Seiberg--Witten}} transform, were achieved by Pal in ~\cite{P}. A boundary state computation was performed by Mukhi in ~\cite{M} for the corrections of order $F^2$ to the Chern--Simons action. Keeping all orders in derivatives in the Seiberg--Witten limit allows to recognize the $\star_2$ product. The purpose of this note is to recover the other predictions. \\

\newpage

 In a boundary state computation, the full Chern--Simons action, including derivative corrections $\Delta S_{CS}$, reads in superspace notation ~\cite{W}~\cite{dVL}, where $\phi^\mu=X^\mu+\theta\psi^\mu$ and $D=\theta\d_\sigma-\d_\theta$ are the superfield and the superderivative: 

$$S_{CS}+\Delta S_{CS}=\langle C|\exp\(-\frac{i}{\l}\int d\sigma d\theta D\phi^\mu A_\mu\)|B\rangle_R\vg$$
  
\noindent{while} the full Dirac--Born--Infeld action reads as follows
{\footnote{The states $|B\rangle_R$ and $|B\rangle_{NS}$ are the boundary states in the Ramond and Neveu--Schwarz sectors with zero field strength.

}}:

$$S_{DBI}+\Delta S_{DBI}= \langle 0|e^{-\frac{i}{\l}\int d\sigma d\theta D\phi^\mu A_\mu(\phi)}|B\rangle_{NS}\pt$$

The Taylor series expansion in terms of the Grassmannian and oscillatory parts of the superfields has to be performed. We will denote by $\tilde{X}^\mu$ and $\tilde{\psi}^\mu$ the non-zero mode parts of ${X}^\mu$ and ${\psi}^\mu$. This expansion leads to two terms. The term containing fermionic zero modes is the one that will yield the corrections to the Chern--Simons action. The other term is alone in the Neveu--Schwarz sector and will therefore yield the corrections to the Dirac--Born--Infeld action:

$$\int d\sigma d\theta D\phi^\mu A_\mu(\phi)=-\int d\sigma d\theta\sum_{k\geq 0}\frac{1}{(k+1)!}\frac{k+1}{k+2}D\tilde{\phi}^\nu \tilde{\phi}^{\mu}\tilde{\phi}^{\mu_1}\dots\tilde{\phi}^{\mu_{k}}\d_{\mu_1}\dots\d_{\mu_{k}}F_{\mu\nu}(x)$$
$$-\int d\sigma(\tilde{\psi}^\mu\psi_0^\nu+{\psi}_0^\mu\psi_0^\nu)\sum_{k\geq 0}\frac{1}{k!}\tilde{X}^{\mu_1}\dots\tilde{X}^{\mu_k}F_{\mu\nu}(x)\pt$$

\section{Boundary state computation: the coupling to $C^{(6)}$}

We first review the computation ~\cite{M} of the coupling to $C^{(6)}$.
 The degrees of the couplings are governed by the number of fermionic zero-modes, since the zero-mode expectation values will be eventually given ~\cite{W} by the substitution

$$\langle C|\frac{1}{2}\psi_0^\mu\psi_0^\nu F_{\mu\nu}|B\rangle_R \mapsto -i\alpha'F\pt$$

 In order to find derivative corrections to the form $C^{(6)}\wedge F^2$, one has to extract terms containing two field strengths and four fermionic zero-modes from the above expansion:

$$\(S_{CS}+\Delta S_{CS}\)|_{(6)}=\frac{1}{2}\sum_{n\geq 0}\sum_{p\geq 0}\(\frac{i}{\l}\)^2\int_0^{2\pi}d\sigma_1\int_0^{2\pi}d\sigma_2\langle C|\(\frac{1}{2}\psi^\mu_0\psi^\nu_0\)\(\frac{1}{2}\psi^\alpha_0\psi^\beta_0\)\times$$
$$\frac{1}{n!}\tilde{X}^{\lambda_1}(\sigma_1)\dots\tilde{X}^{\lambda_n}(\sigma_1)\frac{1}{p!}\tilde{X}^{\rho_1}(\sigma_2)\dots\tilde{X}^{\rho_p}(\sigma_2)\times$$
$$\d_{\lambda_1}\dots\d_{\lambda_n} F_{\mu\nu}(x)\d_{\rho_1}\dots\d_{\rho_p} F_{\alpha\beta}(x)|B\rangle_R\pt$$

The result of a contraction between a pair of $\tilde{X}$'s is obtained by using the properties of the boundary state  $|B(F)\rangle$ to remove annihilation operators from the oscillatory expansion of one of the $\tilde{X}$'s. To evaluate the contractions we need a regularized propagator $D^{\rho\lambda}$. It will be expressed in terms of some specific limit of

$$h^{\mu\nu}=\(\frac{1}{g+\l(B+F)}\)^{\mu\nu}\pt$$

With the regulator used in ~\cite{W} one gets a propagator whose derivative with respect to $\epsilon$ is a geometric sum, so that we have for the regularized expression:

$$D^{\mu\nu}(\epsilon,\sigma_{12})=\alpha'\sum_{m\geq 1}\frac{e^{-\epsilon m}}{m}\(h^{\mu\nu}e^{im(\sigma_2-\sigma_1)}+h^{\nu\mu}e^{-im(\sigma_2-\sigma_1)}\)$$
$$=\alpha'\(h^{\mu\nu}\ln(1-e^{-\epsilon+i\sigma})+h^{\nu\mu}\ln(1-e^{-\epsilon-i\sigma})\)\vg$$

\noindent{where} we denoted the translation-invariant argument by $\sigma_{ij}=\sigma_i-\sigma_j$. We note that the powers of $\alpha'$ will drop out in the Seiberg--Witten limit. In this limit we have the  scaling $g\sim\alpha'^2$ at fixed $B$, so that when we work at fixed order in the field strength, $h^{\mu\nu}$ becomes a constant antisymmetric tensor:

$$h^{\mu\nu}=\frac{\theta ^{\mu\nu}}{\l}\pt$$

We are looking for the regular contributions to the coupling. The contractions between pairs of scalars select the contributions with the same number of scalars at $\sigma_1$ and $\sigma_2$ because contractions between scalars located at the same point contribute an overall $\ln\,\epsilon$ factor. The term in the action containing $C^{(6)}$ therefore reads:

$$\frac{1}{2}\int C^{(6)}\wedge V^{(4)}\vg$$

\noindent{where}

$$V^{(4)}=\sum_{n\geq 0}\frac{1}{n!}\frac{1}{(2\pi)^n}\int_0^{2\pi}\frac{d\sigma_1}{2\pi}\int_0^{2\pi}\frac{d\sigma_2}{2\pi}   \[\ln\(\frac{1-e^{-\epsilon+i(\sigma_1-\sigma_2)}}{1-e^{-\epsilon-i(\sigma_1-\sigma_2)}}\)\]^n\times$$
$$\theta^{\lambda_1\rho_1}\dots\theta^{\lambda_n\rho_n}\d_{\lambda_1}\dots\d_{\lambda_n}F(x)\wedge\d_{\rho_1}\dots\d_{\rho_n}F(x)\pt$$

One of the integrations drops out thanks to periodicity and translation invariance. The principal determination of the complex logarithm $\ln(e^{-i(\sigma-\pi)})$ yields:

$$V^{(4)}=\sum_{n\geq 0}\frac{1}{n!}\(-\frac{i}{2\pi}\)^n\int_0^{2\pi}\frac{d\sigma}{2\pi}(\sigma-\pi)^n\times$$
$$\theta^{\lambda_1\rho_1}\dots\theta^{\lambda_n\rho_n}\d_{\lambda_1}\dots\d_{\lambda_n}F(x)\wedge\d_{\rho_1}\dots\d_{\rho_n}F(x)$$
$$=\sum_{p\geq 0}\frac{(-i)^{2p}}{2^{2p}(2p+1)!} \theta^{\lambda_1\rho_1}\dots\theta^{\lambda_n\rho_n}\d_{\lambda_1}\dots\d_{\lambda_{2p}\rho_{2p}}F(x)\wedge\d_{\rho_1}\dots\d_{\rho_{2p}}F(x) $$
$$=\frac{\sin(\frac{\d_\lambda\theta^{\lambda\rho}\d'_{\rho}}{2})}{\frac{\d_\lambda\theta^{\lambda\rho}\d'_{\rho}}{2}}F(x)\wedge F(x')|_{x'=x}$$
$$=\star_2[F^2(x)]\pt$$

This agrees with the prediction coming from the agreement between the commutative and non-commutative Chern--Simons actions in the Seiberg--Witten limit.

\section{Lower-degree Ramond--Ramond fields and contractions}

In order to generalize the last result to lower-degree Ramond--Ramond form fields, we look for the corrections to the term 

 $$\int C^{(4)}\wedge F^3\pt$$

We are therefore instructed to extract the cubic term in the field strength from $\langle C|B(F)\rangle_R$, together with six fermionic zero-modes, and to investigate which configurations of the scalar oscillators give rise to regular contributions:

$$\(S_{CS}+\Delta S_{CS}\)|_{(4)}=\frac{1}{3!}\sum_{n\geq 0}\sum_{p\geq 0}\sum_{q\geq 0}\(\frac{i}{\l}\)^3\int_0^{2\pi}d\sigma_1\int_0^{2\pi}d\sigma_2\int_0^{2\pi}d\sigma_3$$
$$\langle C|\(\frac{1}{2}\psi^\mu_0\psi^\nu_0\)\(\frac{1}{2}\psi^\alpha_0\psi^\beta_0\)\(\frac{1}{2}\psi^\kappa_0\psi^\tau_0\)\times$$
$$\frac{1}{n!}\tilde{X}^{\lambda_1}(\sigma_1)\dots\tilde{X}^{\lambda_n}(\sigma_1)\frac{1}{p!}\tilde{X}^{\rho_1}(\sigma_2)\dots\tilde{X}^{\rho_p}(\sigma_2)\frac{1}{q!}\tilde{X}^{\phi_1}(\sigma_3)\dots\tilde{X}^{\phi_q}(\sigma_3)\times$$
$$\d_{\lambda_1}\dots\d_{\lambda_n} F_{\mu\nu}(x)\d_{\rho_1}\dots\d_{\rho_p} F_{\alpha\beta}(x)\d_{\phi_1}\dots\d_{\phi_q} F_{\kappa\tau}(x)|B\rangle_R\pt$$

The sum can be reorganized in terms of the number of propagators of each type, say $D^{\mu\nu}(\sigma_{ij})$ appearing after the contractions. The {\it type} means the choice of $i$ and $j$. Let $A$, $B$, $C$ be the numbers of indices of $X(\sigma_1)$ contracted with   $X(\sigma_2)$, of $X(\sigma_2)$ contracted with   $X(\sigma_3)$, of $X(\sigma_3)$ contracted with   $X(\sigma_1)$. As we are looking for regular contributions, self-contractions are forbidden and the regular contributions come from the configurations for which $n=A+C$, $p=A+B$ and $q=B+C$.

$$\(S_{CS}+\Delta S_{CS}\)|_{(4)}=\frac{1}{3!}\sum_{A\geq 0}\sum_{B\geq 0}\sum_{C\geq 0}\(\frac{i}{\l}\)^3\int_0^{2\pi}d\sigma_1\int_0^{2\pi}d\sigma_2\int_0^{2\pi}d\sigma_3$$
$$\langle C|\(\frac{1}{2}\psi^\mu_0\psi^\nu_0\)\(\frac{1}{2}\psi^\alpha_0\psi^\beta_0\)\(\frac{1}{2}\psi^\kappa_0\psi^\tau_0\)\times$$
$$\frac{1}{(C+A)!}\(\tilde{X}^{\lambda_1}\dots\tilde{X}^{\lambda_{C+A}}\)(\sigma_1)\frac{1}{(A+B)!}\(\tilde{X}^{\rho_1}\dots\tilde{X}^{\rho_{A+B}}\)(\sigma_2)\frac{1}{(B+C)!}\(\tilde{X}^{\phi_1}\dots\tilde{X}^{\phi_{B+C}}\)(\sigma_3)\times$$
$$\d_{\lambda_1}\dots\d_{\lambda_{(C+A)}} F_{\mu\nu}(x)\d_{\rho_1}\dots\d_{\rho_{(A+B)}} F_{\alpha\beta}(x)\d_{\phi_1}\dots\d_{\phi_{(B+C)}} F_{\kappa\tau}(x)|B\rangle_R\pt$$

A symmetry factor is to be computed before the integrals are performed. As a warming up we compute it for the cubic field strength corrections. We will next show by induction how to generalize the result to higher powers of the field strength.

  We have three sets of indices (three sets of scalars) to contract together: $\{\lambda\}$, $\{\rho\}$, $\{\phi\}$. Let us first choose $A$ indices from $\{\lambda\}$, to be contracted with indices from $\{\rho\}$. Then, for each of these indices, we choose one index from $\{\rho\}$ and substitute the corresponding two-point function $D^{\lambda\rho}(\sigma_{12})$. We have $C_{A+C}^A\times (A+B)!/B!$ ways of doing it. The $B$ remaining indices from $\{\rho\}$ have to be contracted with indices from $\{\phi\}$; this procedure brings a factor of $(B+C)!/C!$ and the corresponding functions $D^{\rho\phi}(\sigma_{23})$. The $C$ remaining indices from $\{\phi\}$ are then contracted with the remaining ones from $\{\lambda\}$ in $C!$ ways, giving rise to the functions $D^{\phi\lambda}(\sigma_{13})$. The symmetry factor to be inserted into the expression after propagators have been substituted for scalars is obtained as:

$$\frac{(A+C)!}{A!}\frac{(A+B)!}{B!}\frac{(B+C)!}{C!}\,.$$

 The expression is rewritten after the contractions:

$$\(S_{CS}+\Delta S_{CS}\)|_{(4)}=\frac{1}{3!}\(\frac{i}{\l}\)^3\sum_{A\geq 0}\sum_{B\geq 0}\sum_{C\geq 0}\int_0^{2\pi}d\sigma_1\int_0^{2\pi}d\sigma_2\int_0^{2\pi}d\sigma_3$$

$$\frac{1}{A!}\frac{1}{B!}\frac{1}{C!}\(D^{\lambda_1\rho_1}\dots D^{\lambda_A\rho_A}\)(\sigma_{12})\times$$ 
$$\(D^{\rho_{(A+1)}\phi_1}\dots D^{\rho_{(A+B)}\phi_B}\)(\sigma_{23})\times$$
$$\(D^{\phi_{(B+1)}\lambda_{(A+1)}}\dots D^{\phi_{(B+C)}\lambda_{(A+C)}}\)(\sigma_{13})\times$$
$$C^{(4)}\wedge\d_{\lambda_1}\dots\d_{\lambda_{(A+C)}} F\wedge\d_{\rho_1}\dots\d_{\rho_{(A+B)}} F\wedge\d_{\phi_1}\dots\d_{\phi_{(B+C)}} F\pt$$

 For more sets of indices the number of such {\it triangles} $(\{\lambda\}, \{\rho\}, \{\phi\})$ is higher. Consider $K$ sets of indices, corresponding to derivative corrections to $C^{(10-2K)}\wedge F^K$. There are $P_K:=C_K^2$ different propagators involved: $D^{\mu\nu}(\sigma_{ij})$, with $1\leq i<j\leq K$. We will often drop the index $K$ in the sequel, and simply write $P$. A graph with $K$ vertices is associated to each way of performing the contractions, the links being associated to the propagators. The allowed graphs are those where there is no link between any of the vertices and this vertex itself. We claim that the symmetry factor to be inserted into the summand, in front of the propagators and form fields, has the same form as we explicitly computed for $K=3$:

  $$\prod_{i=1}^{P}\frac{1}{N_i !}\vg$$

\noindent{so that}, in a short-hand notation we have at order $F^K$:

$$\(S_{CS}+\Delta S_{CS}\)|_{(10-2K)}=\frac{1}{K!}\sum_{N_1\geq 0}\sum_{N_2\geq 0}\dots\sum_{N_P\geq 0}\(\prod_{i=1}^{P}\frac{1}{N_i !}  \)\(\prod_{i=1}^{P}H_i^{N_i}\) \times$$
$$C\wedge (\partial\dots\partial F)^K\vg$$

\noindent{where} $H_i$ is a propagator of the $i$-th type, carrying a pair of upper indices to be contracted with two lower indices carried by partial derivatives acting on two different $F$'s.\\

{\bf Proof.}  We checked the step $K=3$ (actually it would have been sufficient to take the explicit computation for $K=2$ as a first step). Suppose the claim holds at level $K$, and increase $K$ \mbox{by 1.} This amounts to adding a new set of scalars located at $\sigma_{K+1}$, and to integrating over $\sigma_{K+1}$.
We get $K$ new types of propagators, those connected to the new set of scalars at $\sigma_{K+1}$. They come in numbers $N_{P+1},\dots,N_{P+K}$, which means that $N_{P+i}$ contractions are made between the scalars at $\sigma_{K+1}$ and the scalars at $\sigma_i$, for $1\leq i\leq K$. Each way of contracting the scalars gives rise to a graph with $K+1$ vertices, with a subgraph consisting of the first $K$ vertices and links between them. In this subgraph, let $S_i$ be the number of links connected to the $i$-th vertex, for $1\leq i\leq K$. That is,  for $1\leq i\leq K$, the $i$-th vertex   is connected $N_{K+i}$ times to the vertex labelled $K+1$, and $S_i$ times to vertices in the subgraph. Let also $N_J$, with $J\leq P$, be the number of contractions of each type $J$ inside the subgraph (suppose we have labelled these $P$ types of contractions by integers). The $S_i$'s are functions of the $N_J$'s, the numbers of contractions of each kind within the subgraph, over which there will be a summation. In order to obtain the symmetry factor for the graph with $K+1$ indices, we have to multiply the symmetry factor of the subgraph by the number of ways of making the contractions with the vertex labelled $K+1$. According to the induction hypothesis, this is, in our notations:

$$\sum_{N_{1}}\dots\sum_{N_{P+K}}\left(\(\prod_{j=1}^K\frac{1}{(S_j+N_{P+j})!}\)\times\frac{1}{(N_{P+1}+\dots+N_{P+K})!} \right)\times$$
$$\(C^{N_{P+1}}_{N_{P+1}+\dots +N_{P+K}}C^{N_{P+2}}_{N_{P+2}+\dots+N_{P+K}}\dots C^{N_{P+K-1}}_{N_{P+K-1}+N_{P+K}} \)\times
\prod_{j=1}^K \frac{(S_j+N_{P+j})!}{S_j!}\times$$
$$ \(\frac{\prod_{j=1}^K S_j!}{\prod_{I=1}^P N_I!}\)\pt$$

   The first factor comes from the expansion of the exponential function; it is the product of the factorials of the numbers of indices over the vertices. The second one counts the number of ways of choosing indices over the vertex at $\sigma_{K+1}$ and of doing the contractions between this vertex and the $K$ remaining vertices from the subgraph. The last factor comes from the subgraph with $K$ vertices and the induction hypothesis; we restored in it $\prod_{j=1}^K S_j!$, which was compensated at step $K$ by the factorials arising from the expansion of the exponential function. At step $K$, what we called $S_i$ is just the number of scalars in the set located at $\sigma_i$. What remains inside the summand after simplification is just :

$$\frac{1}{(N_{P+1}+\dots+N_{P+K})!} C^{N_{P+1}}_{N_{P+1}+\dots +N_{P+K}}C^{N_{P+2}}_{N_{P+2}+\dots+N_{P+K}}\dots C^{N_{P+K-1}}_{N_{P+K-1}+N_{P+K}}\frac{1}{\prod_{i=1}^P N_i!}=$$
$$\frac{1}{\prod_{i=1}^P N_i!}\frac{(N_{P+K}+N_{P+K-1})!}{N_{P+K}!N_{P+K-1}!}\frac{(N_{P+K}+N_{P+K-1}+N_{P+K-2})!}{(N_{P+K}+N_{P+K-1})!N_{P+K-2}!}\times$$
$$\dots\frac{(N_{p+K}+N_{p+K-1}+\dots+N_{p+1})!}{\dots N_{p+1}!}\times$$
$$\frac{1}{(N_{P+1}+\dots+N_{P+K})!}=$$

$$\frac{1}{\prod_{i=1}^{P_K+K}N_i!}\pt$$

 As $P_{K+1}=P_K+K$, this proves the claim.\\

\section{Corrections to the Chern--Simons action in the \\ Seiberg--Witten limit}

We come back for a while to the specialization to the correction at order $F^3$ and explicitly evaluate the contractions. We use the same regularized version of the propagator as above, in the Seiberg--Witten limit. We are left with the following expression for the coupling to $C^{(4)}$:

$$ \(S_{CS}+\Delta S_{CS}\)|_{(4)}=\frac{1}{ 3!}\sum_{A,B,C\geq 0}\(\frac{i}{2\pi}\)^{A+B+C}\frac{1}{A!B!C!}\times$$
$$\int_0^{2\pi}\frac{d\sigma_1}{2\pi}\int_0^{2\pi}\frac{d\sigma_2}{2\pi}\int_0^{2\pi}\frac{d\sigma_3}{2\pi}\(\ln\(\frac{1-e^{-\epsilon+i(\sigma_1-\sigma_2)}}{1-e^{-\epsilon-i(\sigma_1-\sigma_2)}}\)\)^A\times$$
$$\(\ln\(\frac{1-e^{-\epsilon+i(\sigma_2-\sigma_3)}}{1-e^{-\epsilon-i(\sigma_2-\sigma_3)}}\)\)^B\(\ln\(\frac{1-e^{-\epsilon+i(\sigma_1-\sigma_3)}}{1-e^{-\epsilon-i(\sigma_1-\sigma_3)}}\)\)^C\times$$
$$\theta^{\lambda_1\mu_1}\dots\theta^{\lambda_A\mu_A}\theta^{\mu_{A+1}\rho_1}\dots\theta^{\mu_{A+B}{\rho_B}} \theta^{\rho_{B+1}\lambda_{A+1}}\dots\theta^{\rho_{B+C}\lambda_{A+C}}\times$$
$$C^{(4)}\wedge\d_{\lambda_1}\dots\d_{\lambda_{A+C}}F\wedge\d_{\mu_1}\dots\d_{\mu_{A+B}}F\wedge \d_{\rho_1}\dots\d_{\rho_{B+C}}F\pt $$

We are going to use the principal determination of the complex logarithm, and to change the variables in the integrals with $\tau_{ij}:=\sigma_{ij}/(2\pi)$. The integral within the action reads

$$\int_0^1 d\tau_1 \int_0^1  d\tau_2\int_0^1  d\tau_3 (\ln[e^{2i\pi(\tau_{12})+i\pi})])^A
 (\ln[e^{2i\pi(\tau_{23})+i\pi}])^B (\ln[e^{2i\pi(\tau_{13})+i\pi}])^C\pt$$

Within the integration domain, if $\tau_{ij}>0$,  then $0<\tau_{ij}<1$ and $-\pi<2\pi\tau_{ij}-\pi<\pi$.  If $\tau_{ij}<0$,  then $-1<\tau_{ij}<0$ and $-\pi<2\pi\tau_{ij}+\pi<\pi$. This yields the following action, where $\epsilon(\tau_{ij})$ now means the sign of $ \tau_{ij}$:

$$ \(S_{CS}+\Delta S_{CS}\)|_{(4)}=\frac{1}{ 3!}\sum_{A,B,C\geq 0}\(\frac{i}{2}\)^{A+B+C}\frac{1}{A!B!C!}\times$$
$$ \int_0^1 d\tau_1 \int_0^1  d\tau_2\int_0^1  d\tau_3  (2\tau_{12}-\epsilon(\tau_{12}))^A (2\tau_{23}-\epsilon(\tau_{23}))^B(2\tau_{13}-\epsilon(\tau_{13}))^C \times$$
$$\theta^{\lambda_1\mu_1}\dots\theta^{\lambda_A\mu_A}\theta^{\mu_{A+1}\rho_1}\dots\theta^{\mu_{A+B}{\rho_B}} \theta^{\rho_{B+1}\lambda_{A+1}}\dots\theta^{\rho_{B+C}\lambda_{A+C}}\times$$
$$\d_{\lambda_1}\dots\d_{\lambda_{A+C}}F(x)\wedge\d_{\mu_1}\dots\d_{\mu_{A+B}}F(x)\wedge \d_{\rho_1}\dots\d_{\rho_{B+C}}F(x)\pt $$

We recognize the sum of integrals as the Fourier transform 

$$\int_0^1 d\tau_1 \int_0^1  d\tau_2\int_0^1  d\tau_3\exp\(-\frac{i}{2}\sum_{1\leq i<j\leq 3} \d_{i,\mu}\theta^{\mu\nu}\d_{j,\nu}(2\tau_{ij}-\epsilon(\tau_{ij}))\)$$

\noindent{of} the integration kernel ~\cite{L} associated to the $\star_3$ product :

$$\star_3[F^3(x)]=\(\prod_{i=1}^3 \int\frac{d^{10} k_i}{(2\pi)^{10}}\)e^{i(k_1+k_2+k_3)x}J_3(k_1,k_2,k_3)\tilde{F}(k_1)\wedge \tilde{F}(k_2)\wedge \tilde{F}(k_3)$$
$$= \int_0^1 d\tau_1 \int_0^1  d\tau_2\int_0^1  d\tau_3\exp\(-\frac{i}{2}\sum_{1\leq i<j\leq 3} \d_{i,\mu}\theta^{\mu\nu}\d_{j,\nu}(2\tau_{ij}-\epsilon(\tau_{ij}))\)F^3(x)
\pt $$

The coupling to $C^{(4)}$ we have just computed is therefore:

$$\frac{1}{3!}C^{(4)}\wedge\star_3[F\wedge F\wedge F]\pt$$

 With $p$ powers of $F$ to be extracted from the initial formula, we would recognize the $p$-ary product $\star_p$, thanks to  
 the inductive computation of the symmetry factor, together with the Seiberg--Witten limit in the evaluation of the two-point function of the scalars. It is encoded in the integration kernel

$$ J_p(k_1,\dots,k_p)=\int_0^1 d\tau_1 \dots \int_0^1 d\tau_p \exp\(-\frac{i}{2}\sum_{1\leq i<j\leq p} k_{i,\mu}\theta^{\mu\nu}k_{j,\nu}(2\tau_{ij}-\epsilon(\tau_{ij}))\)\pt$$

This is sufficient to claim that a boundary state computation confirms the predictions using non-commutativity for the coupling to the Ramond--Ramond form fields:

$$S_{CS}+\Delta S_{CS}=\sum_{p=0}^{5}\frac{1}{p!}\int C^{(10-2p)}\wedge \star_p[F^p]\pt$$

\section{Corrections to the Dirac--Born--Infeld action in the \\
Seiberg--Witten limit}

We are instructed to expand the following expression at all orders in derivatives and at quadratic order in the field strength:

$$\langle 0|e^{-\frac{i}{2\pi\alpha'}\int d\sigma d\theta D\phi^\mu A_\mu(\phi)}|B\rangle_{NS}\pt$$

Use will be made of the regularized propagators, already written in the Seiberg--Witten limit: 

$$G^{\mu\nu}:=D^{\mu\nu}-\theta_1\theta_2 K^{\mu\nu}  \vg$$
$$ D^{\mu\nu}(\sigma_{12}):= \frac{\theta^{\mu\nu}}{2\pi} \sum_{n\geq 1}\frac{e^{-\epsilon n}}{n}(e^{in(\sigma_2-\sigma_1)}-e^{-in(\sigma_2-\sigma_1)}) \vg$$
$$ K^{\mu\nu}(\sigma_{12}):=\frac{i\theta^{\mu\nu}}{2\pi}\sum_{r\in{\bf Z}+\frac{1}{2}}e^{-\epsilon r}(e^{ir(\sigma_2-\sigma_1)}+ e^{-ir(\sigma_2-\sigma_1)} )\pt $$

In the Neveu--Schwarz sector the zero-mode of the superfield is just a scalar and we have to compute the corrections coming from the oscillatory part of the superfield:

$$D\phi^\mu A_\mu(\phi)=\sum_{k\geq 0}\frac{1}{(k+1)!}D\tilde{\phi}^\mu\tilde{\phi}^{\lambda_1}\dots\tilde{\phi}^{\lambda_{k+1}}\d_{\lambda_1}\dots\d_{\lambda_{k+1}}A_\mu(x)\pt $$

Thus we are led to the following integral expression, from which we will extract the corrected \mbox{Dirac--Born--Infeld action} at order $F^2$ in the Seiberg--Witten limit :

$$\langle 0|\int d\sigma_1 d\theta_1 \int d\sigma_2 d\theta_2 \sum_{k\geq 0}\frac{1}{(k+1)!}D\tilde{\phi}^\mu(\sigma_1)\tilde{\phi}^{\mu_1}\dots\tilde{\phi}^{\mu_{k+1}}(\sigma_1)\d_{\mu_1}\dots\d_{\mu_{k+1}}A_\mu(x)\times$$
$$ \sum_{p\geq 0}\frac{1}{(p+1)!}D\tilde{\phi}^\nu(\sigma_2)\tilde{\phi}^{\nu_1}\dots\tilde{\phi}^{\nu_{p+1}}(\sigma_2)\d_{\nu_1}\dots\d_{\nu_{p+1}}A_\nu(x)|B\rangle_{NS}\pt$$

This is going to be evaluated without contraction between two derivatives, which can be achieved by integrating by part over superspace.\\

We make the following statement about the Grassmann coordinates: the terms to be retained in the integral over superspace are those that arise from two contractions between a derivative $D\tilde{\phi}$ and a superfield $\tilde{\phi}$, because these terms come with the factor $\theta_1\theta_2$. From the following expression

$$D_1 G_{12}^{\mu\nu}=\theta_1\d_1 D_{12}^{\mu\nu}-\theta_2 K_{12}^{\mu\nu}$$

\noindent{and the fact that $D_1 G_{12}^{\mu\nu}=-D_2 G_{12}^{\mu\nu}$, one gets convinced by the statement.}\\

Once the two derivatives have been contracted, the remaining superfields have to be contracted using only the first term $D^{\mu\nu}(\sigma)$ in the propagator, since we are now looking for terms with no Grassmannian part any more. The terms with $k=p$ give the finite contribution to the above expression, and each of them gives rise to $k$ powers of $\theta$ with indices contracted with derivatives acting on the potentials. The two remaining derivatives acting on each of the potentials will be contracted with the indices involving two-point functions of the form $D\tilde{\phi}(\sigma_i)\tilde{\phi}(\sigma_j)$. Before going into further detail, we could note that the consistency between the predictions using non-commutativity and the first orders in derivatives in the boundary state computation, as checked in ~\cite{DMS}~\cite{P}, together with the above statement, could already convince us that the effect of taking all orders in derivatives into account will just complete the $\star_2$ products. We know that we have the right overall coefficient and we see what the series looks like.\\

Let us now specify these two-point functions.  They can be of two kinds, since $i=j$ and $i\neq j$ both yield finite parts. We will call the ones for which $i=j$ inner contractions and the ones for which $i\neq j$ outer contractions. Since we have to deal with two groups of superfields, localized at $\sigma_1$ and $\sigma_2$, we get in each term either two inner contractions ($i$) or two outer contractions $(ii)$. An inner contraction results in the appearance of a propagator of the form denoted by $G^{\mu\nu}|_{1\rightarrow 2}$.\\

$(i)$ {\bf Terms with inner contractions.} This has to be understood under overall integrations, and as a matrix element between $\langle 0|$ and $|B\rangle_{NS}$:

$$\sum_{k\geq 0}\frac{1}{k!^2}\(\(\sum_{i=1}^{k+1}D_1G^{\mu\mu_i}|_{1\rightarrow 2} \prod_{1\leq j\leq k+1,j\not{=}i}\tilde{\phi}^{\mu_j}\)\times \(\sum_{l=1}^{k+1}D_2G^{\nu\nu_l}|_{1\rightarrow 2} \prod_{1\leq m\leq k+1, m\not{=}l}\tilde{\phi}^{\mu_m}\)\)\pt$$

We thus connected the derivative at $\sigma_1$ to the superfield $\tilde{\phi}^{\mu_i}(\sigma_1)$ (and  the derivative at $\sigma_2$ to the superfield $\tilde{\phi}^{\mu_l}(\sigma_2)$). From the ways of choosing $\mu_i$ and $\mu_l$, we will obtain $(k+1)^2$ multiplying the summand. Now we have to connect the two groups of remaining superfields between each other, which procedure is the same as in the computation of the Chern--Simons coupling at order $F^2$, bringing a symmetry factor $k!$, leading to the explicit form

$$\frac{i^2}{2!} \sum_{k\geq 0}\frac{1}{k!}\int_0^{2\pi} d\sigma_1\int_0^{2\pi} d\sigma_2  \int d\theta_1 \int d\theta_2 \;D_1G_{12}^{\mu\alpha_1}|_{1\rightarrow 2} D_2G_{12}^{\nu\beta_1}|_{1\rightarrow 2}\times$$
$$ D^{\alpha_2\beta_2}(\sigma_{12})\dots  D^{\alpha_{k+1}\beta_{k+1}}(\sigma_{12})\d_{\alpha_1}\d_{\alpha_2}\dots\d_{\alpha_{k+1}}A_\mu(x)\d_{\beta_1}\d_{\beta_2}\dots\d_{\beta_{k+1}}A_\nu(x)\pt   $$

Let us compute what will be the contribution of the two derivatives of $G$ after regularization:

$$D_1G_{12}^{\mu\alpha_1}|_{1\rightarrow 2} D_2G_{12}^{\nu\beta_1}|_{1\rightarrow 2}\rightarrow   -\theta_1\theta_2(2i)^2\[\frac{\theta^{\mu\a_1}\theta^{\nu\b_1}}{(2\pi)^2}\(\sum_{n\geq 1}e^{-\epsilon n}-\sum_{r\in{\bf Z}+\frac{1}{2}}e^{-\epsilon r}\) ^2 \]\pt$$

As $\sum_{n\geq 1}e^{-\epsilon n}-\sum_{r\in{\bf Z}+\frac{1}{2}}e^{-\epsilon r}=-1/2+O(\epsilon)$, the contribution of the two inner contractions is as follows:

$$D_1G_{12}^{\mu\alpha_1}|_{1\rightarrow 2} D_2G_{12}^{\nu\beta_1}|_{1\rightarrow 2}\rightarrow   -\theta_1\theta_2 \frac{\theta^{\mu\a_1}\theta^{\nu\b_1}}{(2\pi)^2}\pt$$

After integration over the Grassmannian coordinates we obtain the integral expression of the contribution $(i)$ to the action. The factors of $2\pi$ from the previous expression are used to normalize the integrals over $\sigma_1$ and  $\sigma_2$:

$$\(S_{DBI}+\Delta S_{DBI}\)|_{(i)}=-\frac{i^2}{2!}\int_0^{2\pi}\frac{d\sigma_1}{2\pi}\int_0^{2\pi}\frac{d\sigma_2}{2\pi}\sum_{k\geq 0}\frac{1}{k!}\(\frac{-i}{2\pi}\)^k(\sigma_{12}-\pi)^k\times$$
$$\theta^{\mu\alpha_1}\theta^{\nu\beta_1}\theta^{\a_2\b_2}\dots\theta^{\a_{k+1}\b_{k+1}}\d_{\alpha_1}\dots\d_{\alpha_{k+1}} A_{\mu}(x)\d_{\beta_1}\dots\d_{\beta_{k+1}} A_{\nu}(x)$$
$$=-\frac{1}{8}\theta^{\mu\alpha_1}\theta^{\nu\beta_1}\sum_{p\geq 0}\frac{(-i)^{2p}}{2^{2p}(2p+1)!}\theta^{\alpha_2\b_2}\dots\theta^{\alpha_{2p+1}\b_{2p+1}}\d_{\alpha_2}\dots\d_{\alpha_{2p+1}}F_{\mu\alpha_1}\d_{\beta_2}\dots\d_{\beta_{2p+1}}F_{\nu\beta_1}$$
$$=-\frac{1}{8}\theta^{\mu\alpha_1}\theta^{\nu\b_1}\star_2[F_{\mu\alpha_1},F_{\nu\b_1}]\pt$$

Use of the antisymmetry of $\theta$ has been made to express the result in terms of field strengths.\\

$(ii)$ {\bf Terms with outer contractions.} We first consider a subset of these terms, those in which the derivatives are not contracted with each other. For fixed $k$ we have $(k+1)^2\times k!$ ways of generating such terms, from which we obtain:

$$\frac{i^2}{2!}\sum_{k\geq 0}\frac{1}{k!}\int_0^{2\pi}d\sigma_1\int_0^{2\pi}d\sigma_2\int d\theta_1\int d\theta_2 D_1 G_{12}^{\mu\nu_1}D_2G_{12}^{\mu_1\nu} \times$$
$$D_{12}^{\m_1\nu_1}\dots D_{12}^{\m_{k+1}\nu_{k+1}}\d_{\mu_1}\dots\d_{\mu_{k+1}}A_\mu(x)\d_{\nu_1}\dots\d_{\nu_{k+1}}A_\nu(x)\pt$$

The integration over Grassmannian variables changes the contribution of the derivatives of the superfields as follows:

$$ D_1 G_{12}^{\mu\nu_1}D_2G_{12}^{\mu_1\nu}\rightarrow  -\(\frac{i}{2\pi}\)^2\theta^{\mu\nu_1}\theta^{\mu_1\nu} \vg $$

\noindent{so} that the contribution of the first subset of terms with outer contractions reads

$$ -  \frac{1}{2}\sum_{k\geq 0}\frac{1}{k!}\int_0^{2\pi}\frac{d\sigma_1}{2\pi}\int_0^{2\pi}\frac{d\sigma_2}{2\pi}\(\frac{-i}{2\pi}\)^k(\sigma_{12}-\pi)^k\times$$
$$ \theta^{\mu\nu_1}\theta^{\mu_1\nu}\theta^{\mu_2\n_2}\dots\theta^{\mu_{k+1}\n_{k+1}}\d_{\m_1}\dots\d_{\m_{k+1}}A_\m(x)\d_{\n_1}\dots\d_{\n_{k+1}}A_\n(x) \pt$$

Now we have to take into account the possible contraction between $D\tilde{\phi}^\mu$ and $D\tilde{\phi}^\nu$. In order to parallel the form of the terms just computed, we prevent these contractions from showing up by integrating by part, as was announced, finding a factor of $-1$. For fixed $k$ we get $k+1$ terms (instead of the one with a contraction between the derivatives), in each of which $D\tilde{\phi}^\mu$ is contracted with $\tilde{\phi}^\nu$, since we kept the index structure in the process. We also obtain a factor of $k!$ from the numbers of ways of forming pairs of indices from the two sets $\{\m_1,\dots,\m_{k+1}\}$ and $\{\n_1,\dots,\n_{k+1}\}$. The above computation then leads to the following contribution from this second subset of terms:

 $$\frac{1}{2}\sum_{k\geq 0}\frac{1}{k!}\int_0^{2\pi}\frac{d\sigma_1}{2\pi}\int_0^{2\pi}\frac{d\sigma_2}{2\pi}\(\frac{-i}{2\pi}\)^k(\sigma_{12}-\pi)^k \times$$
$$\theta^{\mu\nu}\theta^{\mu_1\n_1}\dots\theta^{\m_{k+1}\n_{k+1}}\d_{\m_1}\dots\d_{\m_{k+1}}A_\m(x)\d_{\n_1}\dots\d_{\n_{k+1}}A_\n(x) \pt$$

 The sum of these two subsets of terms can be expressed in terms of field strengths since $\theta$ is antisymmetric:

$$\frac{1}{2}\(-\theta^{\mu\nu_1}\theta^{\mu_1\nu}+\theta^{\mu\nu}\theta^{\mu_1\n_1}\)\d_{\m_1}A_\m \d_{\n_1}A_\n= \frac{1}{8}\(-\theta^{\mu\nu_1}\theta^{\mu_1\nu}+\theta^{\mu\nu}\theta^{\mu_1\n_1}\)F_{\m_1\m}F_{\n_1\n}$$
$$=-\frac{1}{4}\theta^{\mu_1\nu}\theta^{\mu\nu_1}F_{\m_1\m}F_{\n_1\n}\pt$$

The $\star_2$ product is again recognized, as in the computation of the terms in $(i)$.

$$\(S_{DBI}+\Delta S_{DBI}\)|_{(ii)}=\frac{1}{4}\theta^{\m\n}\theta^{\rho\tau}\star_2[F_{\n\rho},F_{\tau\m}]\pt$$

Apart from an overall integration factor, we arrive at the following expression for the corrected Dirac--Born--Infeld action in the Seiberg--Witten limit:

$$S_{DBI}+\Delta S_{DBI}=\frac{1}{4}\theta^{\m\n}\theta^{\rho\tau}\star_2[F_{\n\rho},F_{\tau\m}]-\frac{1}{8}\theta^{\m\n}\theta^{\rho\tau}\star_2[F_{\m\n},F_{\rho\tau}]\pt$$

\section{Conclusion}

In this note we have shown through explicit computations that the geometric predictions from {\mbox{non-commutativity}} actually yield the correct result for the derivative corrections to the {\mbox{Chern--Simons}} and {\mbox{Dirac--Born--Infeld}} actions. This completes the consistency check ~\cite{M} for the coupling to $C^{(6)}$.\\

 The technical interest of the boundary state computation is to yield the whole series of derivative corrections. Obtaining the whole series allows to check the consistency of the predictions with the Seiberg--Witten map, because this map is expressed in terms of an integral evaluated with the smearing prescription along a Wilson line, and this prescription carries the whole $\star_p$ products. With no reference at all to the non-commutative description (even if the attempt was induced by the considerable simplification of the propagator in the Seiberg--Witten limit), a string computation has been shown to confirm $\hat{S}_{CS}+\hat{S}_{DBI}$ as the right candidate for the non-commutative action.\\

  This strengthens our confidence in the chain of geometric arguments from non-commutative field theory. One way of using this chain is to state that the agreement between commutative and non-commutative actions constrains the structure of the commutative action, once we have a serious gauge-invariant candidate for the non-commutative action with constant field strength. The other way is to look for explicit proofs of these geometric constraints, which serve as consistency checks for the duality between commutative and non-commutative descriptions. \\   

 Using the boundary state formalism to recover predictions from non-commutativity also confirms the anticipation by Chu and Ho in ~\cite{CH}, according to which non-commutativity could arise either by imposing boundary conditions as operator constraints, as they did, or by considering boundary conditions as constraints on states, which is the boundary state approach. The results of direct boundary state computations agree with their claim, consistently with the duality between operators and states.\\

\noindent{{{ \Large \bf Acknowledgements}}}\\ 

I would like to thank L. Alvarez-Gaum\'e, J.L.F. Barb\'on, R. Minasian and P. Vanhove for discussions and comments, and D.M. Bueno Monge for useful correspondence.\\

\end{document}